\documentclass[aps,superscriptaddress,twocolumn,floatfix,letterpaper,nofootinbib]{revtex4-1}

\usepackage{nicefrac}
\usepackage{braket}
\usepackage{amsmath, amssymb, graphicx}

\newcommand\toZ[1]{\lfloor #1 \rceil}

\newcommand\cM{\mathcal M}
\newcommand\cB{\mathcal B}

\newcommand\R{\mathbb R}
\newcommand\Z{\mathbb Z}

\newcommand\T{\mathcal{T}}
\newcommand{\E}{\mathcal E}

\newcommand{\mcl}{\mathcal{L}}

\newcommand\beq{\begin{equation}}
\newcommand\eeq{\end{equation}}

\usepackage[dvipsnames]{xcolor}

\begin{document}

\begin{titlepage}

\title{
Topological Mott Insulators and Discontinuous $U(1)$ $\theta$-Terms
}

\author{Michael A. DeMarco}
\affiliation{Department of Physics, Massachusetts Institute of
Technology, Cambridge, MA 02139, USA}
\email{demarco@mit.edu}

\author{Ethan Lake}
\affiliation{Department of Physics, Massachusetts Institute of
Technology, Cambridge, MA 02139, USA}

\author{Xiao-Gang Wen}
\affiliation{Department of Physics, Massachusetts Institute of
Technology, Cambridge, MA 02139, USA}

\begin{abstract} 
We introduce a lattice field theory that describes the transition between a superfluid (SF) and a bosonic topological Mott Insulator (tMI) -- a $U(1)$ symmetry protected topological phase labeled by an integer level $k$ and possessing an even integer $2k\frac{e^2}{h}$ quantized Hall conductance. 
Our model differs from the usual $2+1$d XY model by a topological term that vanishes on closed manifolds and in the absence of an applied gauge field, which implies that the critical exponents of the SF-tMI transition are identical to those of the well-studied $2+1$d XY transition.
Our formalism predicts a ``level-shift'' symmetry: in the absence of an applied gauge field, the bulk correlation functions of all local operators are identical for any two models whose values of $k$ differ by an integer. 
% differing by the topological term.
%, for example, near the SF-MI and SF-tMI transitions, and hence the extremely well-studied critical exponents of the $2+1$d XY model apply to the SF-tMI transition. 
In the presence of a background gauge field, the topological term leads to a quantized Hall response in the tMI phase, and we argue that this quantized Hall effect persists in the vicinity of the phase transition into the SF phase. Our formalism paves the way for other exact lattice descriptions of symmetry-protected-topological (SPT) phases, and mappings of critical exponents between transitions from symmetry-breaking to trivial states and transitions from symmetry-breaking to SPT states. A similar ``level shift'' symmetry should appear between all group cohomology SPT states protected by the same symmetry.

\end{abstract}


\maketitle

\end{titlepage}

%{\small \setcounter{tocdepth}{1} \tableofcontents }

\noindent
\section{Introduction}\label{sec:Introduction}

The phase transitions that occur in lattice boson systems between Mott insulators \cite{Mott_1949} and superfluids (SF) have been extensively studied \cite{SFMI_Greiner, 2010Sci...329..547B, 2010Sci...329..523D, 2013CRPhy..14..712P,  PhysRevLett.97.060403, PhysRevLett.98.080404, 2010Sci...329..547B}, in part because they embody the competition between kinetic energy and interactions which underlies much of condensed matter physics. If the kinetic energy 
dominates, the bosons condense into a superfluid. On the other hand, repulsive interactions tend to favor the bosons being localized in real space; if they dominate the system forms a Mott Insulator. 

In a seminal work \cite{PhysRevB.40.546}, Fisher et. al. showed that the superfluid-Mott insulator (SF-MI) quantum phase transition generically falls into one of two universality classes. 
In the Mott insulating phase, the average density is pinned to a fixed integer; if the the chemical potential is varied, then the extra bosons (or holes) may condense into a SF, leading to an `ideal' mean-field transition with dynamical critical exponent $z=2$. On the other hand, if the chemical potential is chosen so that particle-hole symmetry is maintained, then the SF-MI transition is actually a multicritical point and lies in the XY universality class, with $z=1$. The XY transition in particular in $2+1$d is in the same universality class as the condensation of superfluid helium at a finite temperature, and so has seen impressive theoretical \cite{Campostrini:2000iw, Gottlob:1993zd, Hasenbusch:1999cc, Guida:1998bx} and experimental \cite{PhysRevLett.76.944, PhysRevLett.84.4894} study.

This story becomes more complex in systems where time-reversal symmetry is broken. In $2+1$d, the insulating state can develop a quantized Hall conductance, 
% leading to the recently discovered $2+1$d
with such states being known variously as $U(1)$ symmetry protected topological phases \cite{CGL1314},
topological Mott Insulators (tMIs), or the Bosonic Integer Quantum Hall Effect \cite{LV1219,SL1301, GERAEDTS2013288}. 
%\xgw{This work seems for fermions. Mott insulator is only for bosons. We should restricted to boson systems and cite works for bosons, such as  } 
These models generated significant excitement; recently a fermionic analogue \cite{2008PhRvL.100o6401R} has been proposed as the mechanism behind the Quantum Anomalous Hall state in twisted bilayer graphene \cite{2020arXiv201107602C} (though we will focus on bosonic tMIs in this paper).
The topological Mott insulating states are labeled by their Hall conductance, which for the bosonic systems we consider is always an even integer multiple of $e^2/h$, viz. $\sigma_{xy} = (2k) \frac{e^2}{h}$ for $k\in\mathbb{Z}$. 

% Given the theoretical discovery of the tMIs, we also have a superfluid-topological Mott insulator (SF-tMI) quantum phase transition. 
As for the SF-MI transition, we can ask about the nature of the phase transition that occurs between a SF and a tMI. 
What should the the critical exponents of this transition be? 
Due to the strong charge density and current fluctuations, the answer is not immediately clear. 

In this work, we will focus on the particle-hole symmetric $z=1$ multicritical point, where we will be able to write down a lattice field theory well-suited for describing the phase diagram at fixed integer boson density. We show that the critical exponents of the 
% (topological) 
SF-tMI $XY$ transition are exactly the same as the regular
% (trivial) 
SF-MI $XY$ transition. Moreover, we will see that the bulk dynamics of all local excitations are identical at or even away from the critical point, unless there is an applied background gauge field. We do this by constructing a well defined lattice model and showing that the bulk dynamics are invariant under a ``level-shift symmetry'' induced by adding a 
% $2\pi$-quantized 
topological $\theta$-term, that changes the Hall conductance by $2k \frac{e^2}{h}$, hence connecting the dynamics of the tMIs to the $k=0$ trivial MI. As our model is well-regulated on the lattice, we can show that this level-shift symmetry is exact, i.e. valid at all relevant energy scales. Crucially, this means that the extraordinary numerical, theoretical, and experimental study of the $2+1$d XY transition is applicable to the SF-tMI transition as well. 

In the presence of a background gauge field, level-shift symmetry is broken and the topological term leads to a Chern-Simons response and quantized Hall conductance. In particular, the topological term causes the vortices which proliferate in the insulating phase to carry charge. We discuss how charged vortices lead directly to the quantized Hall response. This quantized Hall response characterizes the tMI phase and we argue that it persists in the vicinity of the SF-tMI transition.

We would like to stress that for two transitions related by the level-shift symmetry ({\it i.e.} for two models differ only by the topological term), the correlations of any local operators are identical, at and away from the transition point, even though the two transition points have different Hall conductance. This is possible since the level-shift symmetry, {\it i.e.} the topological term, changes the definition of current operators.

The models described in this paper also represent an advance on a purely theoretical front. It has been known for some time \cite{CGL1314,Chen1604, PhysRevB.95.205142} that topological terms for $d+1$-dimensional spacetime lattice models are labeled by elements of the group cohomology $H^{d+1}(G, U(1))$. The cocycles of these models provide actions that are lattice analogs of continuum $2\pi$-quantized topological $\theta$-terms ({\it i.e.} with $\theta=0$ mod $2\pi$), and some of our analysis parallels continuum work \cite{Xu:2011sj}. It has also been known that in order to obtain a nontrivial group cohomology class for continuous groups, one must consider discontinuous cocycles. However, the first explicit expression of these discontinuous cocycles was only found very recently; the $U(1)$ models discovered in \cite{DeMarco:2021erp} and suitably generalized here are the first examples. They follow considerable work placing quantum Hall physics on lattices \cite{Chen:2019mjw, Sun:2015hla}, and join similar works aiming to describe $2+1$d systems with \cite{Wang:2021smv} or without \cite{Bauer:2021fvc} gappable boundaries. They are one of several approaches demonstrating ways around \cite{2021arXiv210702817H, DeMarco:2021erp} a related no-go theorem \cite{2019CMaPh.373..763K, 2021arXiv210710316Z} preventing Hall conductance on the lattice, which does not hold in our case due to the infinite-dimensional on-site $U(1)$ rotor Hilbert space. These discontinuous cocycles are the key to understanding the SF-tMI transition, and also pave the way for studying more complicated transitions involving nonabelian Lie groups.

This paper is laid out as follows. In Section \ref{sec:ModelOverview}, we lay out the lattice model, discuss the level-shift symmetry, and sketch a phase diagram. In particular, the level-shift symmetry immediately shows that all the critical exponents of the SF-MI transition carry over to the SF-tMI transition. Following this, Section \ref{sec:IntegerPhases} discusses the tMI phases in detail, discusses the physical mechanisms behind their Hall conductance, and discusses the physical picture of the SF-tMI transition in terms of proliferation of composite particle-vortex excitations. Following this, we go further in section \ref{sec:halfIntegralPhases}, where we examine the character of the tMI-tMI phase transitions and map them to a model of interacting fermions.

\section{Model Overview}\label{sec:ModelOverview}
We will be concerned with lattice systems of bosons in $2+1d$ at fixed average integer density.
As in the traditional $XY$ model, the field variables in our theory are $U(1)$ rotors living on the sites $i$ of a three-dimensional Euclidean spacetime lattice. We denote the field variables by $\phi_i$, which correspond to the phase of the microscopic boson operator on site $i$ (as we are working at fixed average density, we are allowed to work solely with the phase modes $\phi_i$). For convenience, we will break with convention and let the $\phi_i$ be periodic under shifts by unity, rather than by $2\pi$. Hence a $U(1)$ angle $\phi_i$ will take values in $[0, 1)$, while a group element is given by $g_i = e^{2\pi i \phi_i}$. This will avoid numerous factors of $2\pi$ below, and one may always convert back to the usual notation by replacing $\phi \to \tilde\phi/2\pi$.

Because we are working with angular variables $\phi_i$, and not the rotor variables $g_i = e^{2\pi i \phi_i}$, we must impose a gauge redundancy on all physical quantities. In order to to ensure that the degrees actually remain rotors, we require a ``rotor redundancy'', with all physical quantities being invariant under the replacement
\begin{equation}
\phi_i \to \phi_i + n_i \label{eq:gaugeredun}
\end{equation}
Put more simply, we require that all physical quantities must be periodic functions of $\phi_i$, and will require that the path integral measure not sum over field configurations related by (\ref{eq:gaugeredun}). Beyond this rotor redundancy, our models will also possess two global symmetries. The first is the $U(1)$ symmetry of boson number conservation, which acts as 
\begin{equation}
\phi_i \to \phi_i + \theta\label{eq:U1sym}
\end{equation} 
%which sends $e^{2\pi i \phi_i}\to e^{2\pi i \theta}e^{2\pi i \phi_i}$ for $\theta$ constant over spacetime. 
where $\theta$ is a constant.
We also have a charge conjugation symmetry:
\begin{equation}
\phi_i \to -\phi_i, \label{eq:CCsym}
\end{equation}
which sends $g_i \to g_i^*$. 
%As usual, there is a qualitative difference in these three relations: eqs. (\ref{eq:U1sym}) and (\ref{eq:CCsym}) are symmetries and can be broken, while eq. (\ref{eq:gaugeredun}) is a redundancy which must never break. 

The trivial (i.e. non-topological) $XY$ model Lagrangian satisfies all of these symmetries. It is
\begin{equation}
S = -\frac{1}{g} \sum_{\braket{i,j}} \cos( 2\pi  (d\phi)_{ij})\label{eq:XYLag}
\end{equation}
Here the sum runs over all links $\braket{i, j}$ in the $2+1$d lattice, and $(d\phi)_{ij} = \phi_i - \phi_j$. The phase of the model is controlled by $g$. As $g\to 0$, the strong `kinetic' term sets $d\phi = 0$ and so $\phi = \text{const.}$, confining all vortices and resulting in the $U(1)$ symmetry breaking superfluid phase. As $g\to\infty$, the fluctuations of $\phi$ overpower the kinetic suppression and vortices proliferate, destroying long-range order and leading to the Mott insulating phase. 

To get the tMIs from a lattice model, we must add a term to the trivial $XY$ action which imparts the model with nontrivial Hall conductance when coupled to a background field. To define this term, we must make use of machinery from simplicial homology, which we briefly describe here. For full details, see the supplemental material of \cite{PhysRevLett.126.021603}. A field which assigns a variable to each zero-dimensional lattice site, such as our $\phi_i$, is called a 0-cochain. A field which assigns a variable to all the $m$-dimensional subsets of the lattice (links, plaquettes, etc), is called an $m$-cochain. The field $\phi_i$ assigns an $\R/\Z$ variable to each zero-dimensional point of the lattice and so is a zero-cochain. An action, which must assign a real number to the three-dimensional tetrahedra of our lattice, is a 3-cochain. 

Our task is to construct a 3-cochain from the 0-cochain $\phi_i$ which will provide a term that may be added to $S$ in order to give a model with a TMI ground state at large $g$. To bridge the gap between the 0-cochains and 3-cochains, we use the cup product, which takes an $m$-cochain $a_m$ and an $n$-cochain $b_n$ to an $m+n$ cochain $a_m \cup b_n$ (we will often abbreviate $a_m\cup b_n$ as $a_mb_n$), and the lattice differential $d$, which takes an $m$-cochain $a_m$ to an $m+1$ cochain $da_{m+1}$ and satisfies $d^2=0$. On the lattice, $(d\phi)_{ij} = \phi_i - \phi_j$. We will also make use of the Hodge star operator ``$\star$'', which re-interprets an $m$-cochain as a $d+1-m$-cochain on the dual lattice and in the present setting satisfies $\star^2 =  1$. 

As discussed in Section \ref{sec:Introduction}, the topological term will be a discontinuous, but periodic, function of the field variables $\phi_i$. Let $\toZ{x}$ represent the nearest integer to $x$. Then $x - \toZ{x}\in (-\frac{1}{2}, \frac{1}{2}]$ is a suitable function, and physical quantities should take the form $f(\phi - \toZ{\phi}, d\phi - \toZ{d\phi},...)$. The $XY$ model Lagrangian (\ref{eq:XYLag}) is a trivial example of such a function, as the cosine term may be rewritten as $\cos(2\pi (d\phi - \toZ{d\phi}))$.

The term that imparts nonzero Hall conductance to the tMIs is 
\begin{equation}
S_{k} = -2\pi i k\int (d\phi - \toZ{d\phi}) d (d\phi - \toZ{d\phi}),\label{eq:Sk}
\end{equation}
where the coefficient of $-2\pi k$ has been chosen for later convenience. Here, ``$\int$" is shorthand for evaluation against a generator of the top cohomology of $H_*(X)$, and we have omitted all cup products. As $S_k$ is constructed purely out of cup products, $S_k$ is a topological term. We may simplify \eqref{eq:Sk} to:
\begin{equation}
S_k = 2\pi i k\int(d\phi - \toZ{d\phi})  d  \toZ{d\phi}
\end{equation}
as $d^2 = 0$. The full model combines the kinetic energy with the theta term:
\begin{multline}
S_{g, k}[\phi] = -\frac{1}{g}\sum_{\text{links}}\cos (2\pi d\phi) 
+2\pi i k \int (d\phi - \toZ{d\phi})d\toZ{d\phi}\label{eq:action}
\end{multline}
and the complete partition function is:
\begin{multline}
Z_{k, g} = \int D\phi e^{ \frac{1}{g}\sum_{\text{links}}\cos 2\pi d\phi- 2\pi i k \int (d\phi - \toZ{d\phi})d\toZ{d\phi}}
\end{multline}
where the measure $\int D\phi = \left[\prod_{i}\int_{-\frac{1}{2}}^{\frac{1}{2}}d \phi_i \right]$ has already gauge-fixed the rotor redundancy (\ref{eq:gaugeredun}).

We may immediately see that the action (\ref{eq:action}) is invariant under several symmetries. Most importantly, the action is invariant under the rotor redundancy (\ref{eq:gaugeredun}), because $d\phi - \toZ{d\phi}$ is invariant and $d\toZ{d(\phi + n)} = d\toZ{d\phi} + d^2n = d\toZ{d\phi}$. In addition to the rotor redundancy, this action enjoys the $U(1)$ symmetry $\phi \to \phi + \theta$ (\ref{eq:U1sym}) and charge conjugation symmetry $\phi_i \to - \phi_i$ (\ref{eq:CCsym}). An unusual time-reversal symmetry will be discussed later.

\begin{figure}
\begin{center}
\includegraphics[width = .65\columnwidth]{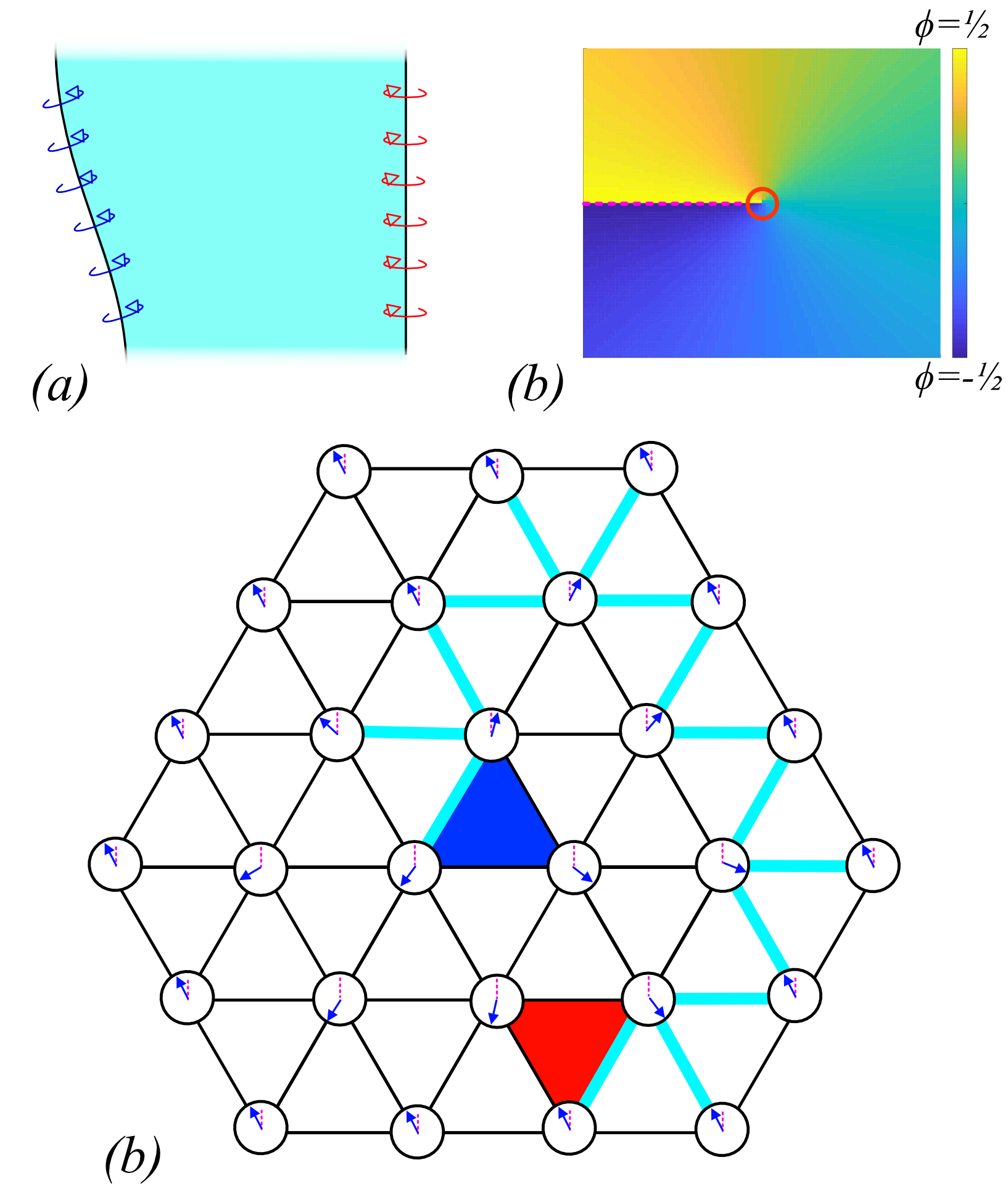}
\caption{\emph{(color online). (a)} In $2+1$ dimensions, $\toZ{d\phi}$ is dual to a surface where $\phi$ has a branch cut, and $\rho_v = -d\toZ{d\phi}$ is dual to the line where a branch cut ends, i.e. a vortex \emph{(b)}. \emph{(c)} Working in two dimensions for convenience, we can see on a microscopic level that  $-d\toZ{d\phi}$ is the vortex number on a plaquette. The field variables are shown as clocks, while links with nonzero $\toZ{d\phi}$ are marked in turquoise. Plaquettes with $-d\toZ{d\phi} \neq 0$ are marked in blue (vortex) and red (anti-vortex).}\label{fig:vortex}
\end{center}
\end{figure}

The key to understanding the topological term is to note that $\jmath_v\equiv \star(- d\toZ{d\phi})$ is the vortex current. To see this, first note that $\toZ{d\phi}$  is a one-cochain which, in three dimensions, is dual to a surface where $\phi$ has a branch cut. Hence $d\toZ{d\phi}$ is dual to the line where the branch cut surface ends (Figure \ref{fig:vortex}a). Branch cut surfaces end at vortex lines; a $2$d slice of a vortex line and branch cut are shown in Figure \ref{fig:vortex}b, where we see that the branch cut terminates in a vortex. On a more microscopic level, consider the $2$d lattice in figure \ref{fig:vortex}c. The $U(1)$ variables on sites are shown as clocks with zero marked as a vertical line. The branch cut links with $\toZ{d\phi} = 0$ are marked in turquoise, and the branch cut terminates in a vortex (blue) and anti-vortex (red).  Hence we see that $- d\toZ{d\phi}$ is indeed the vortex number on a plaquette, with the sign fixed so that $- d\toZ{d\phi} = +1$ at a plaquette hosting a vortex. Note that conservation of vortex number means that $d^\dagger d \jmath_v = 0$, where $d^\dagger \equiv \star d \star$.

In terms of this vortex current $\jmath_v$, we may rewrite the action as:
\begin{equation}
    S_{g, k}[\phi] = -\frac{1}{g}\sum_{\text{links}}\cos( 2\pi d\phi )- 2\pi i k \int (d\phi - \toZ{d\phi})\star \jmath_v,
\end{equation}
where we see that the effect of the topological term is to couple $\phi$ to the vortex current. Moreover, the topological term actually gives the vortices mutual statistics. Written in terms of $\jmath_v$, the topological term on a closed manifold is
\begin{equation}
    S_{k}[\phi] =   -2\pi i k \int  \toZ{d\phi}\star \jmath_v = 2\pi i k  \int \jmath_v \cup  \frac{d}{\square }\jmath_v,
\end{equation}
where $\square$ is the Laplacian. That this term gives statistical phases to braided vortex lines can be seen by noting that it is of the same form as the response of a Chern-Simons gauge field coupled to a background current. More directly, current conservation $d^\dagger \jmath_v = 0$ allows us to write $\jmath_v = \star dL$, where the Poincare dual of $dL$ marks the vortex lines in spacetime. In terms of $L$, 
\beq 2\pi i k \int \jmath_v\cup \frac{d}{\square}\jmath_v = 2\pi i k \int L \cup dL,\eeq 
which indeed gives $2\pi i k$ times the linking number of the vortex worldlines.
Hence we see that the topological term changes the statistics of vortices, with vortex lines having mutual statistics of $e^{2\pi i k}$. 

In the superfluid phase, vortices are suppressed, and we expect that the topological term should be unimportant, leading to just a single SF phase. On the other hand, vortices proliferate in a disordered phase, and so we expect the topological term to be important in the disordered phase. As we will soon see, the tMI fixed points are given by $g\to \infty$, $k\in \mathbb Z$. For generic $k$ and large $g$, we will see that the system flows to attractive fixed points at $\toZ{k}$ and $g\to \infty$ which are studied in Section \ref{sec:IntegerPhases}. For large $g$ and half-odd-integer $k$ we have the tMI-tMI transition discussed in Section \ref{sec:halfIntegralPhases}. 

Given that the topological term charges the statistics of vortices to be $e^{2\pi i k}$, we expect that the bulk dynamics are only sensitive to the value of $k$ mod 1. This is indeed true, and this fact is responsible for constraining the critical exponents of the SF-tMI transition to be equal to those of the usual SF-MI transition. To see this, note that
\begin{equation}
    S_{g, k+1}[\phi] - S_{g, k}[\phi] = 2\pi i \int\Big(d\phi d\toZ{d\phi} - \toZ{d\phi} d\toZ{d\phi}\Big)
\end{equation}
The first term is a surface term, while the second is an integer multiple of $2\pi i$ and may be dropped. Hence, away from a boundary, the bulk dynamics at $k$ and $k+1$ are identical. All correlation functions of local operators, and hence all critical exponents, are identical (see Appendix \ref{app:LS} for a physical interpretation). We will call the bulk invariance under $k\to k+1$ the ``level shift symmetry''.

Continuum topological $\theta$ models have similar level shift symmetries. From a theoretical perspective, shifting the level in this model corresponds to adding an SPT order. In this particular case, it is remarkable that doing so changes the action by only a surface term and therefore does not change the bulk dynamics of the system for any local operator and hence leads to identical critical behavior of local operators at the SF-MI and SF-tMI critical points. More generally, this may be true not just for the SF-tMI transition but for generic symmetry breaking transitions. If two topological phases have the same symmetry, differ only by the addition of an SPT or invertible topological order, and undergo symmetry-breaking phase transitions to the same symmetry-breaking phase, similar arguments may imply that the critical behavior of the two models is identical --- but more study will be required in this area.  

Returning to the model at hand, the level-shift symmetry 
under $k\to k+1$ has important implications for the action of time-reversal symmetry. 
Note that time-reversal reverses the orientation of the lattice and so exchanges the representative of the top cohomology, effectively changing the sign on the integral in the topological term $S_k[\phi]$, which consequently is odd under time reversal. This implies that if $2k \in \mathbb Z$, then the bulk dynamics are time-reversal symmetric. 
\begin{figure}
    \centering
    \includegraphics[width=.75\columnwidth]{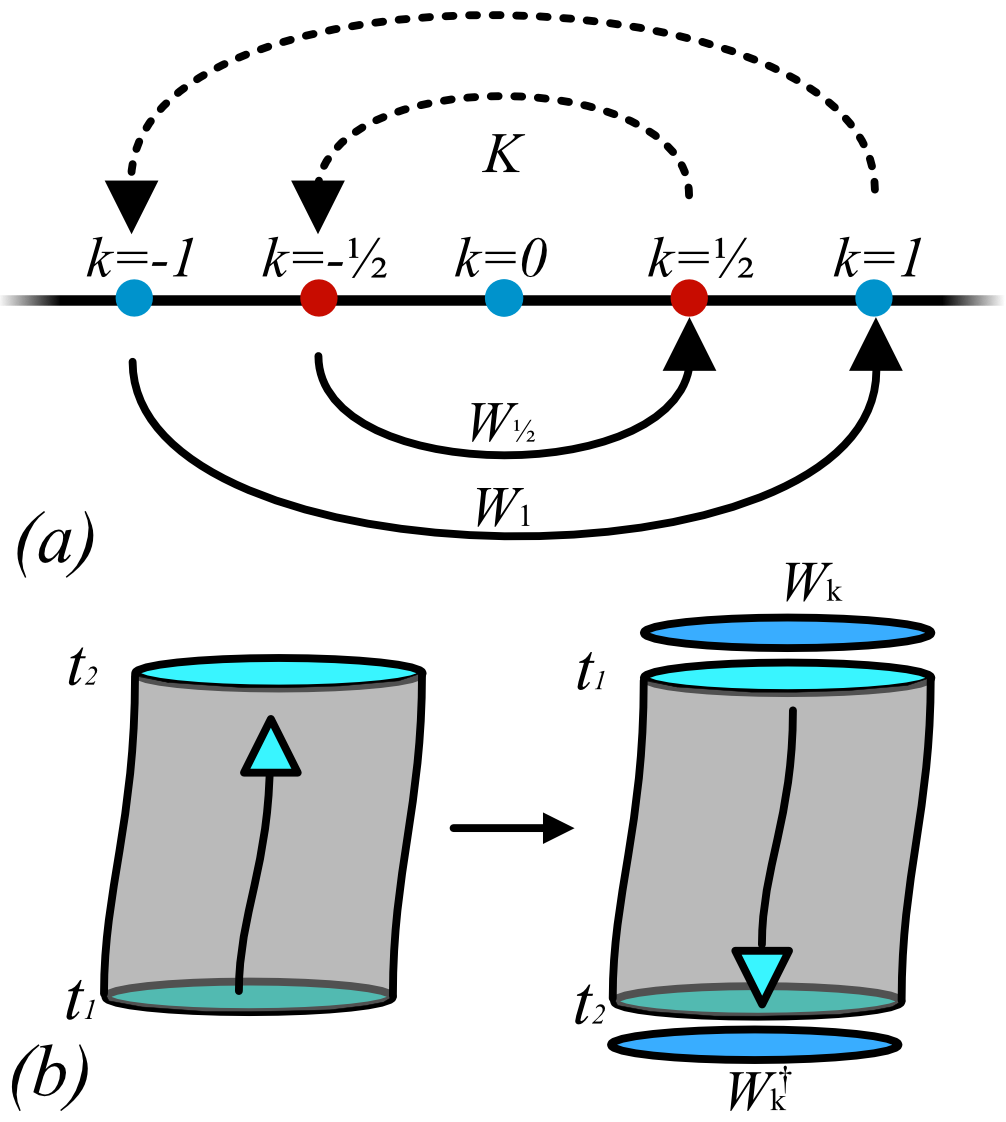}
    \caption{\emph{(color online)} For $2k\in \Z$, the action of the $W_k$ operators cancels out the effects of time-reversal on the boundary of an open manifold.}
    \label{fig:time_reversal}
\end{figure}

While the standard time reversal symmetry does not hold on the boundary, a modified time-reversal symmetry does. Time-reversal symmetry holds for $2k \in \Z$ because $S_{g, k}$ and $S_{g, -k}$ differ only by a surface term:
\begin{equation}
S_{g, k} - S_{g, -k} = 2\pi i (2k) \int d(\phi d\toZ{d\phi})
\end{equation}
On a closed manifold, this surface term vanishes, and the theory at $k$ and its time-reversed conjugate at $-k$ are identical. To extend this symmetry to open manifolds, we need to cancel the leftover effects on the boundary (See Figure \ref{fig:time_reversal}). Let us consider a spacetime manifold $\cM^3$ with a spatial boundary $\cB^2$. For $2k\in \mathbb Z$, we define the operators $\hat\E_{2k} = e^{- 2\pi i (2k)\int_{\cB^2} \phi d\toZ{d\phi}}$, and let $\hat K$ be the complex conjugation operator. We define the time-reversing operators:
\begin{equation}
    \T_k = \hat K \hat \E_{2k}
\end{equation}
For $k=0$, this reduces to the usual time-reversal operator. For other integer or half-odd-integer $k$, the operator $\hat \E_k$ corresponds to shifting the level by $2k$. This is precisely the the boundary change of the bulk level under shifting $k$ by an integer. For each $k$ with $2k\in \Z$, the model is invariant under the $\T_k$ operator.

\begin{figure}
    \centering
    \includegraphics[width = \columnwidth]{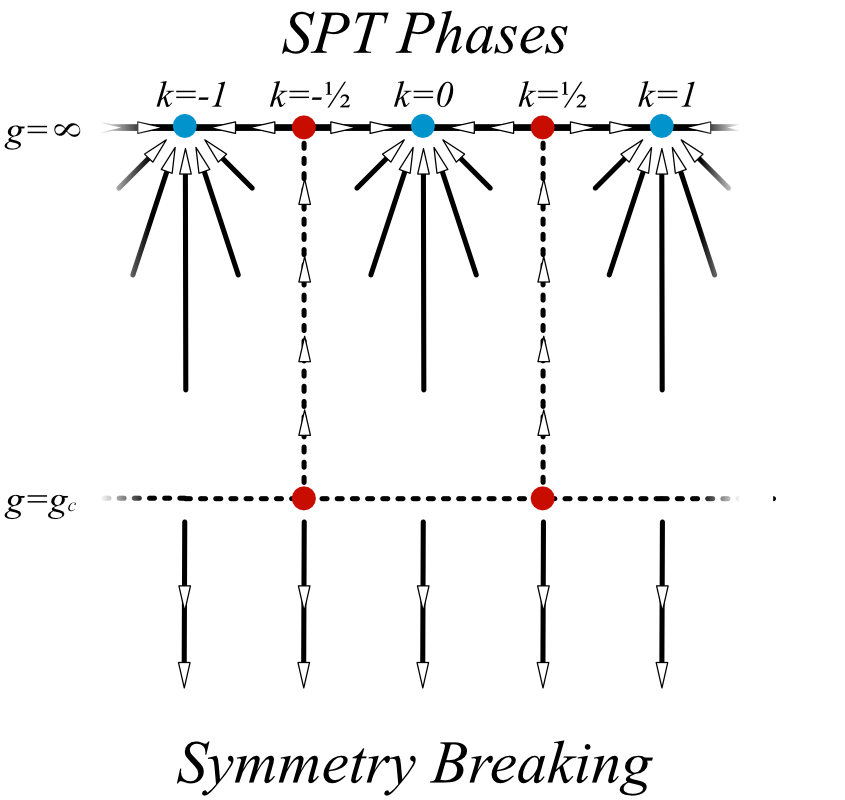}
    \caption{\emph{(color online.)} Because the bulk dynamics are invariant under $k\to k+1$, we may construct a sketch of the RG flow diagram. Along the $g=\infty$ axis are the $\T_k$ fixed points, while at $g=g_c$ we have the SF-(t)MI transitions.}
    \label{fig:Phase_Diagram_Sketch}
\end{figure}

We should combine the fixed lines of $k$ with a rough intuition about the behavior of the model under varying $g$, with $k=0$ fixed. As $g\to 0$, we expect that the model spontaneously breaks $U(1)$ symmetry.
%, setting $\phi = \text{const.}$ 
Conversely, for $g$ greater than a critical $g_c$, vortices proliferate, and the system becomes disordered. Hence we see that, along $k=0$, there should be an unstable fixed point at $g=g_c$, with $g$ either flowing to the symmetry-breaking phase $g=0$ or the MI phase $g\to \infty$. We can then use the level shift symmetry $k\to k+1$ to extend this behavior to all integral $k$. 

Together, these observations lead to the RG flow diagram shown in Figure \ref{fig:Phase_Diagram_Sketch}. The line at $g_c$ separates the disordered phase from the symmetry breaking phase. As there are few vortices in the symmetry breaking phase, we expect that the topological term will be only be relevant in the MI phase. The novel parts of this model are the disordered fixed points with modified time-reversal symmetry with $2k \in \Z$ at $g\to \infty$. Later, we will see that the integer fixed points with $k\in \Z$ describe gapped SPT phases with Hall conductance, while the half-odd-integer fixed points describe the topological transition between tMIs at $k$ and $k+1$.

In the next section, we will study the integer $k$ phases in detail. We will discuss in detail how to see that they describe the topological Mott insulators, and will describe the physical mechanisms behind the SF-tMI phase transition and the tMI Hall conductance. In Section \ref{sec:halfIntegralPhases}, we will study the half-odd-integer phases, and argue that they are the transitions between the tMIs.

\section{The Integral Phases}\label{sec:IntegerPhases}
We now consider the fixed point phases with $g\to\infty$ and $k\in \Z$. In this case, the action reduces to:
\begin{equation}
2\pi i k \int d\phi d\toZ{d\phi} = 2\pi i k \int d(\phi d\toZ{d\phi})
\label{eq:IntAction}
\end{equation}
This is the action of the exactly soluble model in \cite{DeMarco:2021erp}, where it is shown that this action creates a ground state which has a nontrivial Chern number $2k$ and Hall response of $2k \frac{e^2}{h}$. In what follows, we will see that the ground state consists of a condensate of charged vortices, and that this condensation leads to an SPT phase with Hall conductance.

Before proceeding to that ground state physics, it will be useful to examine the model coupled to a background gauge field. In the SF phase (see Appendix \ref{sec:gaugedActions}), we may minimally couple the action to get:
\begin{multline}
    S_{g, k}[\phi; A] = \frac{1}{g}\sum_{\braket{i,j}}\cos 2\pi ((d\phi)_{ij} - A_{ij}) 
    \\
+2\pi i k \int (d\phi - A - \toZ{d\phi-A})d(A + \toZ{d\phi-A})\label{eq:gauged}
\end{multline}
where we have used the form of the action in eq. (\ref{eq:Sk}) to create the minimal coupling. As with $\phi$, we have also taken $A$ to have a period of unity, rather than $2\pi$. Correspondingly we have two rotor redundancies
\begin{align}
    \phi_i \to \phi_i+n_i \\
    A_{ij}\to A_{ij} +m_{ij}
\end{align}
with $m, n \in \Z$, in addition to the usual gauge symmetry:
\begin{align}
    \phi_i \to \phi_i + \varphi_i \nonumber \\ 
    A_{ij} \to A_{ij} + \varphi_i - \varphi_j
\end{align}
and global $U(1)$ symmetry. With $k\in \mathbb{Z}$, the topological part of the action in the superfluid phase becomes:
\begin{multline}
    S_{k}[\phi; A] = 
+2\pi i k \int 
\Big((d\phi-A) dA  -\toZ{d\phi - A}dA \\+ (d\phi - A)d\toZ{d\phi - A}\Big),\label{eq:int_gauged}
\end{multline}
where the term $\int \toZ{d\phi-A} d \toZ{d\phi-A}$ has been dropped on account of it being an integer.
With the background field, the vortex current becomes:
\begin{equation}
    \star \jmath_v = d(d\phi - A - \toZ{d\phi - A}) = -(dA + d\toZ{d\phi - A})
\end{equation}
Evaluating (\ref{eq:int_gauged}) on a closed manifold and substituting in the vortex density, the topological part of the action becomes:
\begin{equation}
    S_{k}[\phi; A] = 
    2\pi i k \int \big(AdA + A\cup(\star \jmath_v) + (\star\jmath_v)\cup  A\big)\label{eq:chargedVortices}
\end{equation}
Thus we see a second effect of the topological term: in addition to giving the vortices mutual statistics at non-integer $k$, at integer $k$ it gives vortices a charge of $2k$. In turn, we will see next that it is the condensation of these charged vortices that gives rise to the tMI phase.

% In the tMI phase, these charged vortex composites are condensed. 

% If the vortices are now charged, how then can they condense without breaking $U(1)$ symmetry? The following subsection \ref{sec:IntegerPhases:GroundState} answers this by showing that is actually composite excitations of charged vortex and oppositely charged particle that condense. In subsection \ref{sec:IntegerPhases:HallConductance} we discuss how the Hall conductance is related to the composite particle-vortex condensate, and in subsection \ref{sec:IntegerPhases:SPTOrder} we discuss how all these properties reveal the SPT character of the tMI phase.  

\subsection{Ground State and SPT Order}\label{sec:IntegerPhases:GroundState}

A closer look at the action (\ref{eq:IntAction}) above reveals an apparent paradox. The action would seem to be trivial, as the Lagrangian $d\phi d\toZ{d\phi} = d(\phi d\toZ{d\phi})$ is a total derivative, and yet it has been shown in \cite{DeMarco:2021erp} that the state is a stable, gapped phase with Hall conductance. The resolution to this reflects the SPT nature of the tMI phase. In particular, the term which the Lagrangian is a total derivative of, viz. $\phi d\toZ{d\phi}$, is itself not locally $U(1)$ symmetric. On a closed manifold, one may rewrite the Lagrangian as a total derivative of $-d\phi [d\phi]$, but in this case the rotor redundancy does not hold locally, and the model thus fails to even be well-defined. There is no way to write the Lagrangian total derivative of a term which is locally both $U(1)$ symmetric and rotor redundant. 

Thus the apparent paradox is not a paradox at all. The Lagrangian is indeed a total derivative, but it is not a derivative of anything $U(1)$ symmetric and rotor redundant. If we break $U(1)$ symmetry, then the model is trivial, and the action can be canceled by an allowed total derivative term. On the other hand, so long as $U(1)$ symmetry is preserved (along with rotor redundancy, which is always required) then the action is nontrivial. 

At the same time, the model is easily solved because it is a total derivative. In particular, the bulk dynamics are trivial: if $\partial \cM^3 = 0$, then the action vanishes and the partition function simply becomes unity. However, the ground state of the model, exposed on a spatial boundary of spacetime, contains nontrivial physics.

We can see this behavior directly from the ground state wavefunction, by which we mean the wavefunction created on the boundary. Specifically, we obtain the ground state on a $2$d manifold $\cB^2$ by evaluating the action \eqref{eq:IntAction} on a spacetime $\cM^3$ with (spacelike) boundary $\cB^2 = \partial \cM^3$. Because the action is a surface term, we can immediately write down the $g\rightarrow\infty$ (un-normalized) ground state:
\begin{equation}
\ket{\psi_k}=\int D\phi e^{2\pi i k \int_{\cB^2} \phi d\toZ{d\phi}}\ket{\{\phi\}}
\label{eq:GroundState}
\end{equation}
where $\ket{\{\phi\}} = \otimes_i \ket{\phi_i}_i$ and again where we gauge-fix the measure as $\int D\phi = \prod_{i\in \cB^2} \int_{-\frac{1}{2}}^{\frac{1}{2}} d\phi_i$.

To understand the ground state further, let us first examine the $k=0$ case. In that case, the $g\rightarrow \infty$ ground state is 
\begin{equation}
\ket{\psi_{k=0}}=\int D\phi \ket{\{\phi\}} = \otimes_i \ket{0}_i\label{eq:keq0gs}
\end{equation}
where $\ket{0} = \int_{-\frac{1}{2}}^{\frac{1}{2}} d\phi \ket{\phi}$ is the $U(1)$ symmetric state with eigenvalue $1$. This is the ground state of a trivial Mott insulator. Recall that $\phi_i$ should be thought of as the phase of a particle at site $i$. In the Mott insulating phase, the particle wavepackets do not overlap and their phases fluctuate independently, as in eq. (\ref{eq:keq0gs}) This should be compared to deep in the symmetry breaking phase, where phase rigidity develops and where all the $\phi_i$ are equal in the ground state. For nonzero $k$, the MI wavefunction is `twisted' by the operator
\begin{equation}
    \mathcal{\hat E}_k = e^{2\pi i k \int \hat \phi d\toZ{d\hat \phi}} = e^{-2\pi i k \int \hat \phi \hat \rho_v}
\end{equation}
so that 
\begin{equation}
    \ket{\psi_k} = \mathcal{\hat E}_k \ket{\psi_{k=0}}
\end{equation}
Thus $\mathcal{\hat E}_k$ is the operator that provides the SPT entanglement to twist the trivial MI into a tMI. Note that on a closed manifold, $\hat E_k$ is $U(1)$ invariant, as under $\phi \to \phi+\theta$,
\begin{equation}
    \hat \E_k \to e^{2\pi i k \theta \int \hat \rho_v}\hat \E_k
\end{equation}
On a closed manifold, the total vortex number must vanish, and so $\int \hat \rho_v = 0$. This is however not necessarily true on a manifold with boundary; hence the $U(1)$ symmetry is anomalous and breaks in the presence of a boundary.

We can also see this behavior locally. Let us split the $\hat \E_k$ into operators on each plaquette $\Delta$ that transform the $k=0$ wavefunction into a nontrivial $k$ ground state, namely: 
\begin{equation}
M_k[\Delta] = e^{-2\pi i k \int_{\Delta} \phi \rho_v}
\end{equation}
so that 
\begin{equation}
    \ket{\psi_k} = \prod_\Delta M_k[\Delta] \ket{\psi_{k=0}} = \hat \E_k \ket{\psi_{k=0}}
\end{equation}
On any plaquette $\Delta$, $M_k[\Delta]$ is not $U(1)$ invariant. However, when multiplied together over a closed manifold, the $M_k$ are indeed $U(1)$ invariant, as
\begin{equation}
    \prod_{\Delta\in \cB^2} M_k[\Delta]
    = \mathcal{\hat E}_k
\end{equation}
which, as discussed above, is $U(1)$ invariant. This behavior is how SPTs are often characterized: if we relinquish $U(1)$ symmetry, then a finite set of local unitary transformations (the $M_k[\Delta]$) suffices to transform the trivial ground state $\ket{\psi_{k=0}}$ into a general $\ket{\psi_k}$. On the other hand, if we require $U(1)$ symmetry, then only the nonlocal operator $\mathcal E_k = \prod_{\Delta} M_k[\Delta]$ suffices.

\subsection{Hall Conductance}

\begin{figure}
    \centering
    \includegraphics[width=.5\columnwidth]{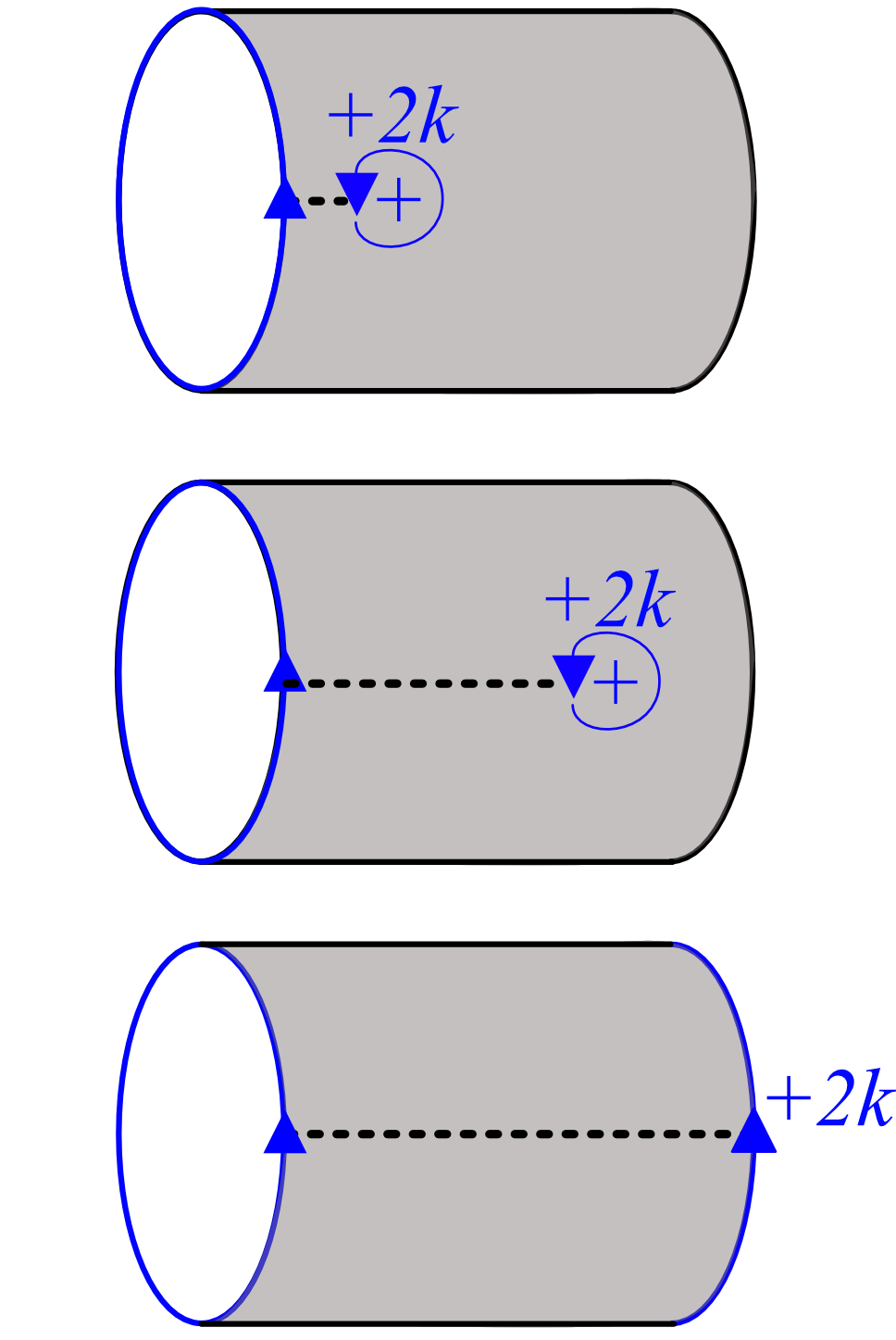}
    \caption{\emph{(color online).} In the Laughlin thought experiment, we twisting the boundary conditions of an open cylinder by a flux quantum transfers charge from one edge to another. We may understand the twisting boundary conditions as the tunneling of a vortex from one edge to another. Because the vortices carry charge $2k$, twisting boundary conditions transfers charge $2k$ from one edge to another and the Hall conductance is $2k$.}
    \label{fig:HallConductance}
\end{figure}

The topological invariant for these phases is the Hall conductance. In \cite{DeMarco:2021erp}, the Chern number of the integer ground states was calculated explicitly by considering the model with twisted boundary conditions, and was found to be $2k$. Physically, we can understand this result in the context of the Laughlin thought experiment. Consider the system on an open cylinder, as shown in Fig. \ref{fig:HallConductance}. We wish to twist the boundary conditions around the open direction by unity. Let $\ket{\psi_k, \theta = 0}$ be the ground state at level $k$ before twisting of boundary conditions and $\ket{\psi_k,\theta = 2\pi}$ be the ground state after an adiabatic twisting process. The two are related by the the creation of a vortex/anti-vortex pair at opposite edges of the sample, with the resulting branch cut providing the $2\pi$ phase jump:
\begin{equation}
    \ket{\psi_k,\theta=2\pi } = \hat V_{\Delta, \Delta'}\ket{\psi_k, \theta = 0},
\end{equation}
where $\hat V_{ij}$ creates a vortex at plaquette $\Delta$ and an anti-vortex at plaquette $\Delta'$. 
We can physically imagine this as dragging a vortex across the cylinder in order to create the branch cut. However, we have seen that the vortices have charge $2k$, and so dragging a vortex across the sample transfers charge from one side of the sample to the other, resulting in a hall conductance of $2k\frac{e^2}{h}$.

This Hall conductance is measurable through the tMI phase, and should persist in the vicinity of the SF-tMI phase transition. Taking into account (\ref{eq:chargedVortices}) near a phase transition, an effective field theory becomes:
\begin{multline}
    S_{\text{eff}}=\alpha(\rho_{SF}) \int (\Pi_T A)^2 + 2\pi i k \int AdA \\
    + \beta(\rho_{SF}) \int (\Pi_T A) d(\Pi_T A) + ...
\end{multline}
where $\Pi_T$ is the transverse projector. The functions
% $\alpha(\rho_{SF}) = 0+\alpha_1 \rho_{SF}+...$ and $\beta(\rho_{SF}) = 0 + \beta_1 \rho_{SF}+ ...$
$\alpha(\rho_{SF}), \beta(\rho_{SF})$
must vanish with $\rho_{SF} \to 0$, since they may only arise from integrating out gapless modes. In particular, as the transition is continuous, no term may suddenly cancel the transverse conductance due to the Chern-Simons term.

We have now examined the tMI integer phases of the model, examined the the entanglement structure of the tMIs, understood the SF-tMI transition as the proliferation charged vortices, and explained how this leads to a Hall conductance in the disordered phase and near the transition. In the next section, we examine the topological transition between phases at $k$ and $k+1$ that occurs for $g\to \infty$.

\section{half-odd-integer Theories}\label{sec:halfIntegralPhases}

We have seen that the integer $k$ theories describe $U(1)$ SPT phases. In turn, we now discuss the half-odd-integer theories, which describe the phase transitions between the SPT phases that occur upon tuning $k$. 

Taking $g\to\infty$ and setting $k = m -\frac{1}{2}$, with $m\in\Z$, the action (\ref{eq:action}) becomes:
\begin{equation}
S_{g\to\infty, k=m - \frac{1}{2}}[\phi] = \pi i (2m - 1) \int_{\cM^3} (d\phi - \toZ{d\phi})d\toZ{d\phi}
\end{equation}
This action is invariant under the time-reversing operator $\T_{m+\frac{1}{2}}$. If $\partial \cM = \emptyset$, then the action reduces to:
\begin{equation}
S_{g\to\infty, k=m - \frac{1}{2}}[\phi] = -\pi i (2m - 1) \int_{\cM^3} \toZ{d\phi}d\toZ{d\phi}. \label{eq:khalfbulk}
\end{equation}

As a first measure, note that we may rewrite the bulk action as the boundary of a time-reversal symmetric term in one higher dimension. For a four-manifold $\mathcal{N}^4$ such that $\partial \mathcal{N}^4 = \cM^3$,
\begin{multline}
    -\pi i (2m - 1) \int_{\cM^3} \toZ{d\phi}d\toZ{d\phi}
    \\
    = 
    -\pi i (2m - 1) \int_{\mathcal N^4} d\toZ{d\phi}d\toZ{d\phi}
\end{multline}
Note that this action, after exponentiation, is invariant under the usual time-reversal. 
For the our purposes, the critical fact is that time-reversal forces $m$ to be an integer in the four dimensional theory, and hence the four dimensional theory is a $U(1)\times \T$ SPT. Our three-dimensional theory, as the boundary of an SPT, must break the $U(1)$ or $\T$ symmetry, be gapless, or develop topological order.

We do not believe that these theories develop topological order. On the one hand, the integer $k$ phases are stable $U(1)$ SPTs; if the half-odd-integer phases were stable topological orders, then there would be to be another class of critical $k$ points, somewhere between the integer and half-odd-integer $k$ (recall Figure \ref{fig:Phase_Diagram_Sketch}). On the other hand, in Appendix \ref{sec:LatticeRG}, we show a numerical RG calculation that suggests that $k$ flows to the nearest integer, with the points where $k$ is a half-odd-integer separating different phases. We find no evidence of a stable phase at half-odd-integer $k$, nor do we find any fixed points besides the integer and half-odd-integer $k$.

Hence we are left with the options of the half-odd-integer phases breaking $U(1)\times \T$ symmetry (in which case the transition between the integer $k$ $U(1)$ SPTs would be first order) or gaplessness (in which case the transition is second order. We now argue that when $k$ is a half-odd-integer, the model admits a description in terms of  emergent fermions, which order in a way dependent on whether or not the transition is continuous.

% In Section [], we will examine the possible RG flow diagrams for this phase across all $k$ and $g$, taking into account the possibilities of first or second order transitions at half integer $k$. However, before examining the full phase diagram, we take the remaining subsections of this section to rewrite the half-odd-integer theory in two suggestive ways: first as a fermion theory with nearest neighbor interactions, and secondly in terms of higher symmetry breaking. 

\subsection{Emergent Fermions}

\begin{figure}
\begin{center}
\includegraphics[width = .75\columnwidth]{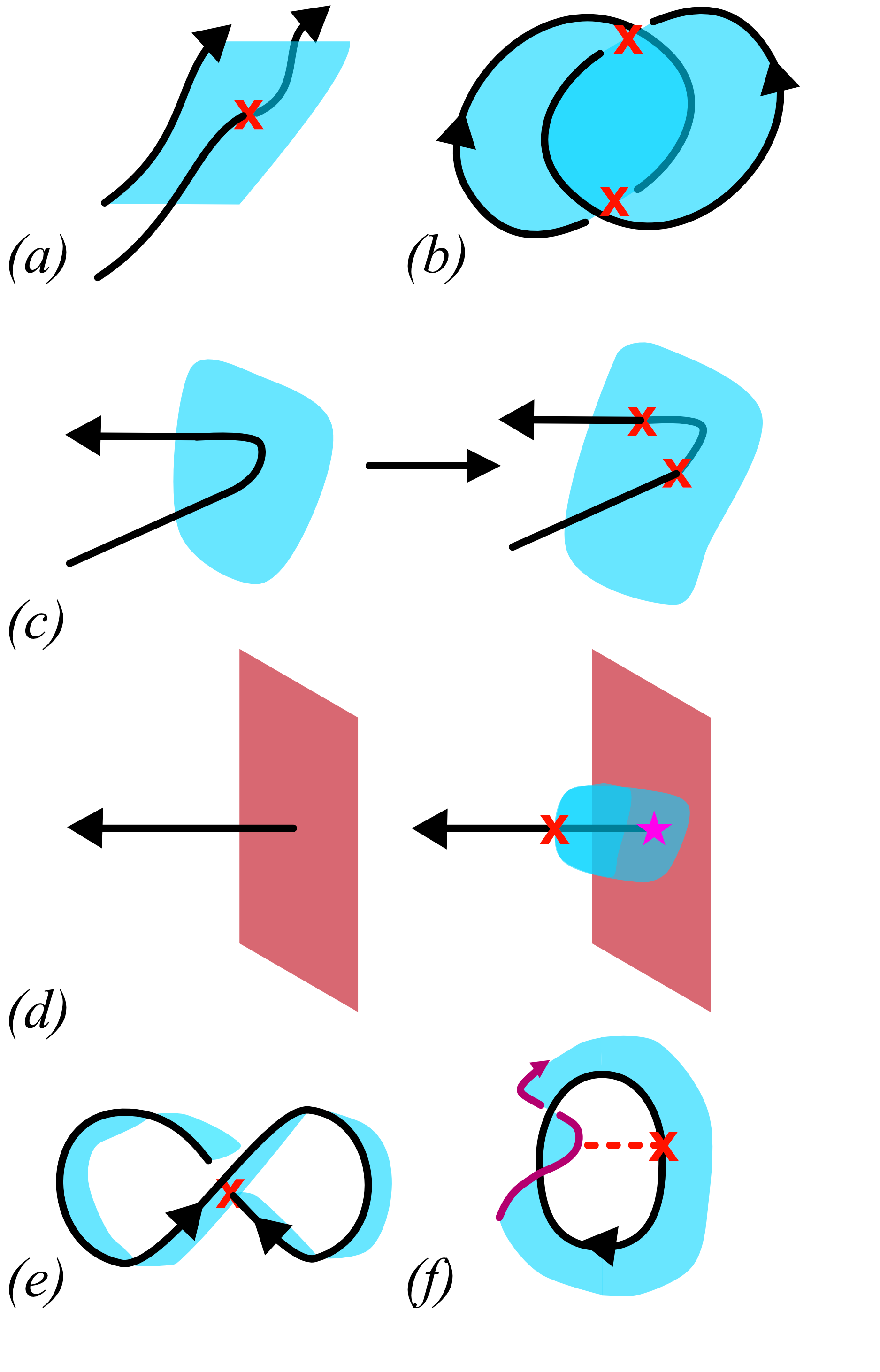}
\caption{\emph{(color online).} $(a)$ The topological term at half integer $k$ assigns a factor of $(-1)$ for each intersection (red $x$) of a vortex line (black) with a branch cut wall (blue). $(b)$ Disconnected lines must have two such intersections and so may not have anyonic statistics. $(c)$ Rotor redundancy $\phi\to \phi+n$ locally move the branch cut walls, and always create intersections in pairs in the bulk. $(d)$ in the presence of a boundary (red), moving a domain wall can create a single intersection, and rotor redundancy is preserved by a charge at the end of the vortex line (magenta star) implemented by the $d\phi$ term.  A single vortex line describing the exchange of two particles has a single intersection, as does the (topologically equivalent) case of $(d)$ a $2\pi$ rotation of a single particle. Hence the vortices are fermions.}\label{fig:fermiStatistics}
\end{center}
\end{figure}

Let us work with the bulk action (\ref{eq:khalfbulk}). Recall that that the vortex current is given by $\jmath_v = -\star d\toZ{d\phi}$. Using this, we can rewrite this action as
\begin{equation}
    -\pi i (2m - 1) \int_{\cM^3} \jmath_v \frac{d}{\Box}\jmath_v
\end{equation}
Recall that in $2+1$d, the $1$-cochain $\toZ{d\phi}$ is dual to a two dimensional surface which describes the branch cut emanating from a vortex. The $2$-cochain $d\toZ{d\phi}$ is dual to a one-dimensional vortex line. The topological $\toZ{d\phi} d\toZ{d\phi}$ term counts the intersections of the vortex lines with the branch cut walls, i.e. twice the linking number (See figures \ref{fig:fermiStatistics}a, \ref{fig:fermiStatistics}b).

In the bulk, this action (\ref{eq:khalfbulk}) is invariant under the rotor redundancy because sending $\phi \to \phi+n$ corresponds to a moving a branch cut wall or creating/annihilating them in pairs. As shown in figure \ref{fig:fermiStatistics}, in the bulk, a local move of a branch cut wall must generate intersections in pairs, hence the action is invariant.

We can further understand the character of these vortices by examining their braiding. First, consider two disconnected, contractible lines. As shown in figure \ref{fig:fermiStatistics}b, these two lines must intersect in two places. In fact $\toZ{d\phi}d\toZ{d\phi}$ is twice their linking number; because the action assigns a factor of $(-1)$ to each intersection, these two lines have no braiding phase, and the vortices do not have anyonic statistics. 

However, a more complex situation arises when we consider just a single vortex line. In figure \ref{fig:fermiStatistics}e, we see a line representing the creation of two vortex-antivortex pairs, the exchange of two vortices, and the annihilation of the swapped pairs. Figure \ref{fig:fermiStatistics}f is topologically equivalent, and shows the case of a $2\pi$ rotation of a vortex. In either case, a single intersection is created, resulting in a factor of $(-1)$. Hence the vortices are fermions.

Now that we have identified the vortices as fermions, it may seem that the task is complete. The content of the action (\ref{eq:khalfbulk}) is exhausted, and a simple guess for the emergent behavior of the system may be completely free fermions. However, this misses a crucial factor. There is a Jacobian we neglected passing from the rotor field $\phi$ to the branch cut and vortex fields $\toZ{d\phi}$ and $\jmath = -\star d\toZ{d\phi}$. This Jacobian leaves a statistical imprint on the theory which will generate interactions for the fermions. In Appendix \ref{app:FermionJacobian}, we show how to map this system to an interacting superconducting system, wherein the $U(1)$ symmetry of our model becomes $S^z$ symmetry. We argue that the order parameter cannot be s-wave. Then the order parameter must break time-reversal symmetry, or be gapless, which corresponds to the symmetry-breaking first-order or gapless situations we have described. It still remains to be determined whether the transition between the integer $k$ SPTs is first or second order.

% In Appendix \ref{app:highersymbreak}, we describe yet another approach to the half integer phases, realizing them by beginning with fermionic SPTs with odd Hall conductance and then explicitly breaking a higher form symmetry. In both of these formulations, it still remains to be determined whether the transition between the integer $k$ SPTs is first or second order.

\section{Discussion}

In this paper, we have used a new discontinuous cocycle model to shed light on the transition between a superfluid and a topological Mott Insulator. Without a background gauge field, the correlation functions of all local operators are identical at and away from the SF-tMI and SF-MI transitions, while introducing a background gauge field leads to different Hall responses for MI and tMI. This immediately implies that all critical exponents of correlation functions of $e^{i2\pi\phi_i}$ and its conjugate momentum must be the same in the SF-MI and SF-tMI phases, which allows the considerable numerical results of the XY model to be applied to the SF-tMI transition. 

We may also ask what other topological phases may be described by a similar framework. An obvious $U(1)\times \T$ phase in $1+1$d is one coming from the action:
\begin{equation}
    S = i\pi k\int_{\cM^2} d\toZ{d\phi}
\end{equation}
On the other hand, $d+1$ dimensional, with even $d>0$, generalizations of our model are given by:
\begin{equation}
    S = 2\pi i k \int_{\cM^{d+1}}(d\phi - \toZ{d\phi})(d\toZ{d\phi})^{\frac{d}{2}}
\end{equation}
where the exponent is taken using the cup product. If, as in $2+1$ dimensions, these are SPT states for integral $k$, they would exhaust all the $U(1)$ SPTs predicted by group cohomology \cite{PhysRevB.87.155114}. Many more possibilities arise with multiple $U(1)$ fields $\phi^I$. One could also consider extending this formalism to SPTs with nonabelian symmetry. Could the Haldane phase be described by a similar exactly solvable discontinuous path integral?
In every case, the existence of a model of the sort we have described here would carry the same implication of level-shift symmetry: in the absence of a gauge field, all correlations of local operators would be the same in the symmetry-breaking to topological transition as in a similar symmetry-breaking to trivial state transition. The SPT order would then by diagnosed by the response to an applied gauge field. 

Even just the model we have described has further applications. Because the edge theory may be treated alone, this proposes a solution to the $U(1)$ ``chiral fermion problem:'' The edge theory must describe pairs of counter-propagating modes with differing charges, so that the total chiral central charge of the edge theory vanishes, while the modes carry a chiral representation of $U(1)$. This will be explored in a future work.

\section{Acknowledgements}
This research is partially supported by NSF DMR-2022428, the NSF Graduate Research Fellowship under Grant No. 1745302, and by the Simons Collaboration on Ultra-Quantum Matter, which is a grant from the Simons Foundation (651440). MD and EL thank Hart Goldman and Jing-Yuan Chen for their insight on Chern-Simons terms. MD and EL thank Victor Albert and Ryan Thorngren for discussions. 

\bibliography{main}

\appendix

\section{Physical Interpretation of Level Shift Symmetry}\label{app:LS}
In the main text, we saw that the fact that the cocycle was a surface term for our model implied that the bulk correlation functions of local operators are identical to the trivial case, and that this implied that the critical exponents for the SF-tMI transition were the same as for the SF-MI transition. Now we discuss why we might expect that on general grounds.

We wish to consider the SF-tMI transition, and we begin in the tMI phase. Rather than directly breaking the symmetry in the tMI, suppose that we stack the tMI with a trivial MI. We can then break the $U(1)$ symmetry in the trivial state, which involves a phase transition with the usual critical exponents. After the symmetry is broken in the trivial phase, we couple in the tMI. With the provision of symmetry breaking, the tMI can be trivialized, and also reduced to a trivial state. The critical exponents for this transition are thus the same as the usual SF-MI transition.

This argument shows that there must always be a SF-tMI critical point with the same exponents as the SF-MI critical point. What we have shown in this paper is that the XY SF-tMI transition, in the absence of a background gauge field, is precisely this transition.

A similar argument may be applied to any SPT, thus showing that there exists a symmetry-breaking transition out of the SPT that has the same exponents as the symmetry breaking transition of a symmetric trivial state. On the mathematical side, this reflects the fact that all group cocycles $\nu$ have the form $\nu = d\mu$, where $\mu$ may not be $G$-symmetric, i.e. that all SPTs are trivial in the absence of symmetry.

\section{Superconductors of Emergent Fermions in the half-odd-integer Phases}\label{app:FermionJacobian}

First, recall that we have defined the path integral measure as
\begin{equation}
\int D\phi = \left[\prod_{i}\int_{-\frac{1}{2}}^{\frac{1}{2}}d \phi_i \right]
\end{equation}
As each $\phi_i \in [-\frac{1}{2}, \frac{1}{2})$, the branch cut field $\toZ{d\phi}_{ij} = \toZ{\phi_i - \phi_j}$ may take values in $\{-1, 0, 1\}$. However, $\toZ{d\phi}$ is not distributed uniformly over these three values. If we let $\phi_i$ and $\phi_j$ be chosen randomly, the probability $p$ that $\toZ{d\phi}$ takes on a given value is 
\begin{equation}
\toZ{d\phi} = \begin{cases}
-1 & p=\nicefrac{1}{8} \\
~0 & p=\nicefrac{3}{4} \\
~1 & p=\nicefrac{1}{8}
\end{cases}
\end{equation}
This immediately leads to a probability for the observation of a vortex. On a given plaquette, the vortex number is given by $\rho_v = - d\toZ{d\phi}$, and on a triangular lattice must take values in $\{-3, -2, ..., 3\}$. One can extrapolate the probabilities from those for $\toZ{d\phi}$ to get:
\begin{equation}
\rho_v=-d\toZ{d\phi} = \begin{cases}
-3 & p=\nicefrac{1}{512}    \\
-2 & p=\nicefrac{18}{512}    \\
-1 & p=\nicefrac{111}{512}    \\
~0 & p=\nicefrac{252}{512}    \\
~1 & p=\nicefrac{111}{512}    \\
~2 & p=\nicefrac{18}{512}    \\
~3 & p=\nicefrac{1}{512}    \\
\end{cases}
\end{equation}
We see that when the $\phi$ are completely random, although having no vortex is more likely than having a vortex with any given vortex charge, having a vortex or anti-vortex is slightly more likely than having none at all.  

The details of the induced interactions are lattice dependent. However, the generic behavior can be described simply as follows. For random $\phi$, approximately one quarter of the lattice plaquettes will host a vortex, while another quarter will host an anti-vortex. Vortices and anti-vortices on adjacent sites have a weak attractive interaction and may annihilate, while both vortices and anti-vortices may hop to adjacent plaquettes with real amplitude. 

Recalling that the vortices and anti-vortices are statistically fermions for half-odd-integer $k$, we may convert this rough theory into an effective fermion theory. We consider a honeycomb lattice (the dual of the triangular lattice) with real hopping coefficients. We denote vortices as spin-up fermions, anti-vortices as spin down, and impose a strong on-site repulsive interaction to prevent overlap. In this context, conservation of vortex current density corresponds to $S^z$ symmetry, while vortex-antivortex annihilation results in superconducting pairing terms. A Hamiltonian for this model is accordingly
\begin{multline}
    H = \sum_{\braket{i,j}} \Big(t\sum_{\sigma}(c^\dagger_{i, \sigma}c_{j\sigma} + c^\dagger_{j, \sigma} c_{i, \sigma}) \\
    + \Delta (c^\dagger_{i\uparrow}c^\dagger{j\downarrow} + c_{j\downarrow}c_{i\uparrow})
    -g c^\dagger_{i, \uparrow}c_{i, \uparrow}c^\dagger_{j, \downarrow}c_{j, \downarrow}
    \Big) 
    \\
    -\mu \sum_{i}\sum_{\sigma}c^\dagger_{i, \sigma}c_{i, \sigma}
    + U\sum_{i}c^\dagger_{i, \uparrow}c_{i, \uparrow}c^\dagger_{i, \downarrow}c_{i, \downarrow}
\end{multline}
Here the $t$ term represents the fermion hopping; $\Delta$ the vortex-antivortex annihilation; $g>0$ the nearest neighbor attractive interactions; $U\to \infty$ the strong on-site repulsion; and $\mu$ the chemical potential. The parameters should be tuned to roughly reproduce the fermion density, hopping, pairing, and attractive interactions from the statistical analysis in Appendix \ref{app:FermionJacobian}. 

Within this framework, we can again see the question of symmetry breaking or gaplessness (or, equivalently, of a first-order or second-order transition) in the topological transition between $k$ and $k+1$ phases. The question is of the form of the pairing wavefunction $\psi_{ij} = \braket{c^\dagger_{i, \uparrow} c^\dagger_{j, \downarrow}}$. The strong on-site repulsion will suppress s-wave pairing, and we are left with p-wave or d-wave pairing. These must either break time-reversal symmetry or be gapless; whichever the dynamics favor will determine the nature of the topological transition.

\section{Gauged Actions}\label{sec:gaugedActions}
The original gauged action for the integer theories was derived in \cite{DeMarco:2021erp} by ungauging a topologically ordered model. In terms of the dynamical gauge field $a$ for the topological order, this was achieved by setting $a = d\phi + A$, where $d\phi$ becomes the variable for the SPT order and $A$ is a background gauge field. The resulting action at $g=\infty$ is:
\begin{multline}
    S = - 2\pi i k \int\Big\{
    (d\phi - A)(dA - \toZ{dA}) - \toZ{dA}(d\phi - A) \\
    - d\big[(d\phi - A)(d\phi - A - \toZ{d\phi - A})\big]\label{eq:ungaugedAction}
    \Big\}
\end{multline}
where we assume that the background field is free of monopoles, allowing us to drop a 1-cup product. Because the rotor model we discuss in this paper is derived by ungauging the topological order, eq. (\ref{eq:ungaugedAction}) must be the correct gauged action, but it is very different from the minimally coupled model (\ref{eq:int_gauged}) discussed in the main text. 

However, the two actions actually coincide in the superfluid phase. Up to one-cup products that encode framing, the actions differ by a factor of:
\begin{multline}
    4\pi i k \int (d\phi - A)(\toZ{dA} + d\toZ{d\phi - A}) \\= 2\pi i k \int (d\phi - A)\toZ{\star j}
\end{multline}
where we have noted that $\toZ{\star j} = \toZ{dA + d\toZ{d\phi - A}} = \toZ{dA} + d\toZ{d\phi - A}$. Hence, in the superfluid phase, when $\toZ{\star j} = 0$, the two actions agree.

\section{Lattice Renormalization Group}\label{sec:LatticeRG}

\begin{figure}
\begin{center}
\includegraphics[width = \columnwidth]{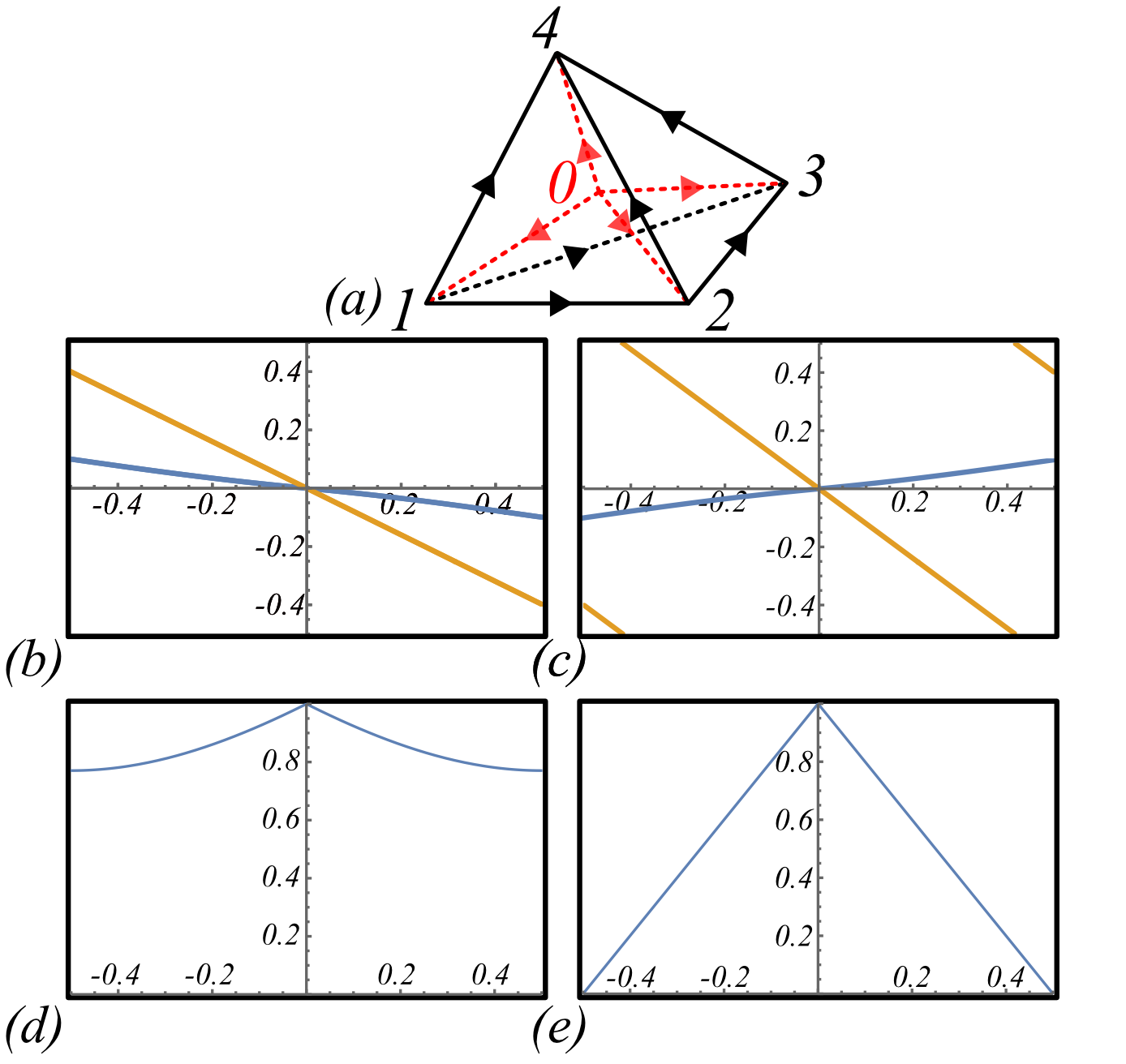}
\caption{\emph{(Color Online.)}}\label{fig:LatticeRG}
\end{center}
\end{figure}

In this appendix we calculate the effects of integrating out a rotor on a given site, with the aim of using the results to build intuition for how $k$ flows under RG. Taking $g\to\infty$ recall that the action in this limit is:
\begin{equation}
    S = 2\pi i k \int d\phi d\toZ{d\phi}.
\end{equation}
To evaluate this, we place it on a tetrahedral lattice shown in Figure \ref{fig:LatticeRG}a and perform a course-graning. Consider integrating out point $0$ in Figure \ref{fig:LatticeRG}a. We wish to calculate:
\begin{widetext}
\begin{multline}
\int d\phi_0 \exp\Bigg\{2\pi i k \Bigg[\left(\phi_1 - \phi_0 - \toZ{\phi_1 - \phi_0}\right)
\Big[e
- \left(\toZ{\phi_2 - \phi_1} + \toZ{\phi_3 - \phi_2} - \toZ{\phi_3 - \phi_1}\right)
\\
+ \left(\toZ{\phi_2 - \phi_1} + \toZ{\phi_4 - \phi_2} - \toZ{\phi_4 - \phi_1}\right)
- \left(\toZ{\phi_3 - \phi_1} + \toZ{\phi_4 - \phi_3} - \toZ{\phi_4 - \phi_1}\right)
\Big] 
\\
+ \left(\phi_2 - \phi_0 - \toZ{\phi_2 - \phi_0}\right)\left[\toZ{\phi_3 - \phi_2} + \toZ{\phi_4 - \phi_5} - \toZ{\phi_4 - \phi_2}\right]
\Bigg]
\Bigg\}
\end{multline}
which can be rewritten as:
\begin{equation}
\int d\phi_0 \exp\Bigg\{-2\pi i k \Bigg[
(\phi_2 - \phi_1 - \toZ{\phi_2 - \phi_0} - \toZ{\phi_0 - \phi_1})
\left(
\toZ{\phi_3 - \phi_2} + \toZ{\phi_4 - \phi_3} - \phi_{4 - \phi_2}
\right)
\Bigg]
\Bigg\}
\end{equation}
Or, setting $\rho_{234} = - \toZ{\phi_3 - \phi_2} + \toZ{\phi_4 - \phi_3} - \toZ{\phi_4 - \phi_2}$, 
\begin{align}
\underbrace{e^{-2\pi i k (\phi_2 - \phi_1 - \toZ{\phi_2 - \phi_1})\rho_v}}_{\Phi_k}
\underbrace{\int d\phi_0 e^{2\pi i k 
( \toZ{\phi_2 - \phi_0} + \toZ{\phi_0 - \phi_1} - \toZ{\phi_2 - \phi_1})
\rho_{234}
}}_{\Phi_c}
\end{align}
Now the amplitude has decomposed into the expected phase $\Phi_k$, times a correction term $\Phi_c$. Turning our attention to the integral in $\Phi_c$, we may use $U(1)$ symmetry to set $\phi_2 = 0$. Then The integral becomes:
\begin{equation}
\Phi_c = \int d\phi_0  e^{2\pi i k (\toZ{\phi_0 - \phi_1} + \toZ{\phi_1} - \toZ{\phi_0})}
=
1-|\phi_1 - \toZ{\phi_1}|+ |\phi_1 - \toZ{\phi_1}| e^{2\pi i k \rho_{234} \text{sgn}(\phi_1 - \toZ{\phi_1})}
\end{equation}
Restoring $U(1)$ symmetry, this is:
\begin{equation}
\Phi_c = 1- |\phi_2 - \phi_1 - \toZ{\phi_2 - \phi_1}|+ |\phi_2 - \phi_1 - \toZ{\phi_2 - \phi_1}| e^{-2\pi i k \rho_{234}\text{sgn}(\phi_2 - \phi_1 -\toZ{\phi_2 - \phi_1})}
\end{equation}
\end{widetext}
To understand this expression, fix $\rho_{234} = 1$. The phase of both $\Phi_k$ and $\Phi_c$ functions of $\phi_2 - \phi_1$ are plotted in figures \ref{fig:LatticeRG}b and \ref{fig:LatticeRG}c for $k = .8$ and $k = 1.2$, respectively. For $k= .8$, the phase of $\Phi_k$ and $\Phi_c$ vary jointly, and $\Phi_c$ is pushing the effective $k$, back towards $k=1$. For $k= 1.2$, they vary oppositely, and $\Phi_c$ is pushing the effective $k$ down, again back towards $k=1$.

This trend continues across the spectrum. For $k$ neither integer nor half integer, $\Phi_c$ acts to push the effective $k$ back towards the nearest integer. For $k\in \Z$, $\Phi_c = 1$, and there is no correction, reflecting the fact that the integral theories are time-reversal-invariant, fixed point theories. For $k\in \Z + \frac{1}{2}$ $\Phi_c$ is real, reflecting the time-reversal invariance of the half-odd-integer theories. 

A further question arises because $|\Phi_c| < 1$ when $k\notin \Z$ and $|\phi_2 - \phi_1 -\toZ{\phi_2 -\phi_1}|>0$. Figures \ref{fig:LatticeRG}d and \ref{fig:LatticeRG}e show $|\Phi_c|$ for $k=.8$ and $k= .5$, respectively. For $ k=.8$, the damping effect is somewhat small, while for $k= .5$ it is extreme, and $|\Phi_c| = 0$ for $|\phi_2 - \phi_1 -\toZ{\phi_2 -\phi_1}| = \pm .5$. For general $k$ and small $|\phi_2 - \phi_1 - \toZ{\phi_2 - \phi_1}|$, this implies that there should be a logarithmic term in the action. Note that this only applies when $\rho_{234} \neq 0$; when $\rho_{234} = 0$, $\Phi_c = 1$ in all cases.

\begin{widetext}

\section{Villain approximations and duality}\label{sec:Villain}

We wish to approximate the action:
\begin{equation}
    S = \frac{1}{2g}\int  |d\phi - A - \toZ{d\phi - A}|^2 + 2\pi i k \int (d\phi - a - \toZ{d\phi-A}) d(d\phi - A -\toZ{d\phi - A})
\end{equation}
by an action which is at most quadratic in the field variables and which does not contain the integer rounding operation $\toZ{\,}$. One way to do this is to use the Villian approximation, which replaces the above with
\begin{equation}
    S_V = \frac{1}{2g}\int |d\phi - A - m|^2 + 2\pi i k \int (d\phi - A - m)d(d\phi - A - m),
\end{equation}
where the integer-valued field $m$ is now summed over in the path integral.
To see the accuracy of this approxmiation, we may shift $m\to m+ \toZ{d\phi - A}$, which yields the action:
\begin{equation}
   S = \frac{1}{2g}\int |d\phi - A - m - \toZ{d\phi - A}|^2 + 2\pi i k \int (d\phi - A - m - \toZ{d\phi - A})d(d\phi - A - m - \toZ{d\phi - A}) \label{eq:villain1}
\end{equation}
Or, setting $\alpha = d\phi - A -\toZ{d\phi - A}$,
\begin{equation}
   S =\frac{1}{2g}\int |\alpha|^2 + 2\pi i k \int \alpha d\alpha  + S_m,\end{equation}
where 
\begin{equation}
S_m = \int\left( -\frac{1}{2g} |m|^2 + \frac{1}{g}\alpha\star m
    \right)- 2\pi i k  \int (m d\alpha + \alpha dm - mdm) 
    \label{eq:villain2}
\end{equation}
The original action can be obtained by setting $m=0$, in which case the error factor vanishes. When $g\to 0$, this is precisely what happens, as $|d\phi - A - \toZ{d\phi - A}|< \frac{1}{2}$ almost always, and hence the strong exponential suppression in eq. (\ref{eq:villain1}) selects $m=0$. 

On the other hand, for $g=\infty$, the error factor in eq. (\ref{eq:villain2}) is invariant under $m\to -m$, and hence is invariant under complex conjugation and must be real. Thus, precisely at $g = \infty$, the error in the Villain approximation cannot affect the topological response.

However, for intermediate $g$, the approximation becomes unreliable. Here we lack the strong exponential suppression, while at the same time the term $\frac{1}{g}(d\phi - A -\toZ{d\phi - A})\star m$ spoils invariance under $m\to - m$, and the error factor is in general complex with a large phase. 

Qualitatively, the behavior of the Villain error term matches what we see in the dual picture; when $g\to 0$, deep in the superfluid phase, the vortices are given mutual statistics and the Chern-Simons response for $AdA$ is intact. This is caused by the strong exponential suppression of $m$ and is in general valid for generic, small $g$. For large $g$, the vortex statistics and Chern-Simons response are again regained, but only at precisely $g= \infty$, where the error term is real. (For large but not infinite $g$, we obtain an extensive error.)
What is interesting here is that the behavior of the dual picture tracks the behavior of the error term in the Villain approximation.

In many applications, the Villain approximation is used outside of its region of strict applicability, on the grounds that the universal behavior of the model in question should be preserved on account of the Villain approximation preserving all symmetries.
However, when we attempt to use the Villain approximation to perform the usual particle-vortex duality mapping on the model considered in the main text, we will see that while the Villain approximation is innocuous in the symmetry breaking phase, it appears to give an incorrect answer for the Hall conductance in the Mott insulating phase. 
% The model we are studying in this paper is the usual 2+1D XY model with the addition of the term proportional to $k$. The physical meaning of this term is that it gives statistical phases to vortex braiding, as can be seen by writing it as 
% \beq 2\pi k \int \jmath^\mu_v d^{-1} \jmath_v^\mu,\eeq 
% where $\jmath_v = \star d(d\phi - \toZ{d\phi})$ is the vortex current. To see that this term is responsible for producing a statistical interaction for the vortices, note that it is precisely of the same form as the response of a $U(1)$ Chern-Simons theory coupled to a background current. 

% In the present description, a statistical interaction like this is necessarily non-local when written in terms of the vortex variables. Since all of the physics that makes this model distinct from the conventional XY model is contained in the statistical vortex interaction, it is convenient to switch to a description in which this physics can be written down in a local way. 

To illustrate this, we attempt to perform particle-vortex duality in the usual way. We start by using the Villain representation of the Lagrangian as given above, and then integrating in a real vector field $X$, giving the Lagrangian (working in continuum notation here and in the rest of what follows)
% as
% \beq \mcl = \frac1{2g}(D_A\phi)^2 + i\bar k D_A\phi \cup d (D_A\phi) + i  d(D_A\phi)\cup B,\eeq 
% where 
% \beq D_A\phi \equiv d\phi -n -A/2\pi,\eeq 
% where $n$ is a dynamical integer-valued 1-cochain, $A$ is a real-valued background field for the global $U(1)$ symmetry,\footnote{The funny $1/2\pi$ factor is due to the fact that our scalar $\phi$ has radius $1/2\pi$.} and $B$ is a real-valued background field which couples to the vortex current. We then integrate in a real-valued vector field $X$ via
\beq \mcl = \frac1{2g} |D_A\phi-X|^2 + i\bar k (D_A\phi- X) \cup d (D_A\phi-X) + i X \cup da + id(D_A\phi-X)\cup B\eeq 
where $\bar k \equiv 2\pi k$,
$D_A\phi \equiv d\phi -n -A/2\pi$, $B$ is a background source for the vortex current, and $a$ is another dynamical real-valued vector field with $da$ having periods quantized in $2\pi \Z$. To see that the addition of the $X$ and $a$ fields hasn't changed anything, we can integrate out $a$: this sets $X$ to both be closed and have integral periods, and as such $X$ can then be absorbed by a shift of $d\phi$ and $n$.

We now shift $X$ by $D_A\phi$. This gets rid of $D_A\phi$ everywhere except in the term $i(D_A\phi)\cup da = i(d\phi-n-A)\cup da$. The first $d\phi \cup da$ part vanishes when integrated over a closed manifold (which we will specify to in what follows), and as such can be dropped. The remaining terms then give 
\beq \mcl = \frac1{2g} |X|^2 + i\bar k X\cup dX + iX \cup da - i (n+A/2\pi)\cup da + idX \cup B.\eeq 
Integrating out $X$ and shifting $a\mapsto a + B$,
\beq \mcl = \frac12 (\star da)_\mu D_X^{\mu\nu} (\star da)_\nu - i (n+A/2\pi)\cup (da+dB),\eeq 
where $D_X^{\mu\nu}$ is the propagator for $X$.
We then softly enforce the constraint coming from summing over $n$ by replacing $n \cup (da+dB)$ with $\lambda \cos(a+B)$ for some positive constant $\lambda$.\footnote{In more precise notation, we replace $\int n\cup d(a+B)$ with $\sum_{l,\mu} \lambda \cos(a_l^\mu+B^\mu_l)$, where the sum runs over all links of the lattice.} We then explicitly separate out the longitudinal part of $a$ by sending $a \rightarrow d\theta + a$, where now $a$ and $\theta$ are connected via $U(1)$ gauge transformations. This gives 
\beq \mcl = \frac12 (\star da)_\mu D_X^{\mu\nu} (\star da)_\nu + \lambda \cos(d\theta +a + B) + i \frac1{2\pi} A \cup (da+dB).\eeq 
 At this point, the only novelty about this theory as compared to the conventional XY model is contained the form of the matrix $D_X$, which in momentum space can be shown to be
 \beq D_X^{\mu\nu}(q)  = \frac{gm^2}{m^2+q^2} \left( \delta^{\mu\nu} + \frac{q^\mu q^\nu}{m^2} - \frac1m \epsilon^{\mu\nu\lambda}q_\lambda \right),\qquad m \equiv \frac 1{2g\bar k}.\eeq 
 This finally gives 
 \beq \mcl = \frac12 \frac{gm^2}{m^2+q^2} \left(|da|^2 +  i\frac{1}{m} a\wedge \square da\right) + \lambda\cos(d\theta+a+B) + i\frac{A}{2\pi} \cup(da + dB),\eeq
 where $\square = -\partial_\mu\partial^\mu$. This representation shows that our original model can be re-written in terms of a gauge theory coupled to matter, with a $\mathcal{T}$-breaking term added to the kinetic term for the gauge field.\footnote{As mentioned above, when $k\in \Z$ the partition function on a manifold without boundary is independent of $k$ in the absence of background fields. Strictly speaking, this is no longer true in the above presentation, due to our use of $\cos(d\theta + a)$ to softly enforce discreteness. In the exact mapping, the $\cos(d\theta+a)$ is replaced by $dn\cup (d\theta+a)$, as written above. Therefore in the absence of background fields, the partition function is 
 \beq Z = \int D n \, \det(D_X D_a)^{-1/2} e^{-\int (\star dn)_\mu D_a^{\mu\nu} (\star dn)_\nu}\eeq
 Now $D_XD_a = (D_X \star d D_X^{-1} \star d) = \star d \star d$, which is independent of $k$. Furthermore, the exponent is just $\int n_\mu [D_X^{-1}]^{\mu\nu}n_\nu$, which for $k\in \Z$ is also independent of $k$. Therefore the exact partition function at $k\in \Z$ is independent of $k$, as it must be.}
 
In the dual representation of the symmetry-breaking phase the gauge field $a$ is not Higgsed, and the term proportional to $\lambda$ can be ignored. The low energy theory is just that of free Maxwell theory, plus irrelevant corrections due to the $a\wedge \square da$ term. Note that the Laplacian appearing in the term $a\wedge \square da$ is essential: without it the gauge field $a$ would be rendered massive, and we would not obtain the correct Meissner effect for $A$ that we know should be present in the symmetry-breaking phase. 

In the dual representation of the disordered phase, $a$ is Higgsed, and picks up a mass term $\frac{M^2}2 a_\mu \Pi_T^{\mu\nu} a_\nu$, where $\Pi_T$ is the transverse projector. Integrating out $a$ using the propagator 
one finds in momentum space (now setting $B = 0$)
\beq \mcl = \frac1{2g}  \frac{m^2+q^2}{m^2(q^2+M^2)^2 + q^6}A_\mu \left( q^2(q^2+M^2) \Pi_T^{\mu\nu} + \frac{q^4}{m} \epsilon^{\mu\nu\lambda}q_\lambda  \right)A_\nu.\eeq 
% In the context of Appendix \ref{sec:Villain}, we now send $g\to\infty$ in order to to avoid errors in the Villain approximation. 
If we set $g=\infty$, we obtain the correct Chern-Simons response, with a Hall conductance of $2ke^2/h$. However, at any finite $g$, the Hall response actually {\it vanishes}, due to the factor of $q^4$ in the $AdA$ term appearing in the above expression. Since the Hall conductance of a gapped phase should not be able to be eliminated by a marginally irrelevant variable (here $1/g^2$), we chalk the vanishing of the Hall conductance at finite but large $g$ to the inaccuracy of the Villain approximation in the large-$g$ limit. 
% \begin{equation}
    % \mathcal L = 2g\bar k \epsilon^{\mu\nu\lambda} A_\mu q_\lambda A_\nu
% \end{equation}
% which is simply the Chern-Simons response. 

% We can also use the dual representation to understand the phase transition which occurs as $g$ is varied with $k$ held fixed. When $k=0$, the transition can famously be described in either representation by the Lagrangians
% \beq \frac{1}{2}(D_A\Phi)^2 + r |\Phi|^2 + \frac{u}{2}|\Phi|^4 + \cdots \leftrightarrow \frac12(D_a\Theta)^2 + u|\Theta|^2 + \frac{v}{2}|\Theta|^4 + \frac{i}{2\pi}a \wedge dA + \cdots,\eeq 
% where $\Phi \sim e^{i\phi}, \Theta \sim e^{i\theta}$. At nonzero $k$, it is helpful to describe the transition starting on the dual gauge theory side discussed above. Here we see that the effect of $k$ is simply to add an additional kinetic term which modifies the dynamics of the gauge field $a$. This term is however less relevant than the Maxwell term for $a$, and as such can be ignored for the purposes of determining the critical theory.

\end{widetext}

\end{document}